\documentclass[]{article}

 \usepackage[]{graphicx}
 \usepackage{epsfig}
 \usepackage{amsmath, amstext, amsfonts}
 \usepackage{amssymb}
 \usepackage{multicol}
 \usepackage[latin1]{inputenc}

\usepackage{anysize} % Soporte para el comando \marginsize
\marginsize{2cm}{2.5cm}{1.5cm}{2cm}

 \title{SOLUTIONS OF THE PERTURBED KDV EQUATION\\
 FOR CONVECTING FLUIDS BY FACTORIZATIONS}
\author{O. Cornejo-P\'erez$^{\dagger}$ and H.C. Rosu$^{\ddagger}$\\
\\
 $^{\dagger}$ Facultad de Ingenier\'{\i}a, Universidad Aut\'onoma de Quer\'etaro\\
Centro Universitario Cerro de las Campanas, 76010 Santiago de Quer\'etaro, Mexico\\
$^{\ddagger}$ Potosinian Institute of Science and Technology,\\
Apdo Postal 3-74 Tangamanga, 78231 San Luis Potos\'{\i}, Mexico\\
\\
 E-mails: octavio.cornejo@uaq.mx, hcr@ipicyt.edu.mx\\
 \\
 arXiv: 0902.0750 v2}

 \date{Central European J. Phys. {\bf 7}, accepted on 26-5-2009}

\begin{document}
%\onecolumn

 \maketitle
% \begin{center}{\sc %% -------------------------------------
% ipicyt %$^{3}$
% \footnotesize \vskip 4ex
%   %$^1$
%   %$^2$
%   Oct-Haret-08.tex\\
%}
% \end{center} %% ----------------------------------------------------
%
\begin{abstract}
  \footnotesize
   \noindent
    In this paper, we obtain some new explicit travelling wave solutions of the perturbed KdV
    equation through recent factorization techniques that can be performed when the coefficients of the equation fulfill a certain condition. The solutions are obtained by using a two-step factorization procedure through which the perturbed KdV
    equation is reduced to a nonlinear second order differential equation, and to some Bernoulli and Abel
    type differential equations whose solutions are expressed in terms
    of the exponential and Weierstrass functions.
%\\
\\

\noindent PACS number(s): 02.30.Hq
 \end{abstract}

 \medskip

%\twocolumn

%%%%%%%%%%% ********************************************************
%%%%%%%%%%% ********************************************************
%%%%%%%%%%% ********************************************************

%% =====================================================================
%\section{Introduction}   \label{S_Int} % 111111
%% =====================================================================

\noindent In a previous paper \cite{fp06}, we have factorized the
Korteweg-de Vries-Burgers (KdVB) equation by means of an efficient
factorization procedure that we introduced in 2005 \cite{rc05}.
This allowed us to obtain in an easy way some travelling wave
solutions of the KdVB equation. Recently, Wang and Li \cite{wl08}
discussed an extended form of our method and applying it to
various nonlinear equations. Our goal in the present paper is to
use jointly the two methods for yet another nonlinear evolution
equation, the so-called perturbed Korteweg de Vries (PKdV)
equation, which is one of the most general nonlinear equations
with important applications \cite{ad90}. We shall use the
following form of this equation
%..................
\begin{equation}\label{can0}
u_t+\lambda(u_{xxx}+6uu_x)+5\beta uu_x+(u_{xxx}+6uu_x)_x=0~.
\end{equation}
It was used to describe the evolution of long surface waves in a
convecting fluid. It has been thoroughly investigated by Cerver\'o
and Zurr\'on in 1996 \cite{cz96}. We notice here that the solution
of a slightly more complicated equation describing the evolution
of a system exhibiting an oscillatory instability with respect to
the static state can be obtained from $u$ with an appropriate
scaling followed by a constant shift proportional to the excess of
the Rayleigh number above its critical value (see page 5 in
\cite{cz96}).

\medskip

Passing to the travelling variable $z=x-ct$, we convert this
equation after integrating it once into the following ODE
%..................
\begin{equation}\label{can00}
u_{zzz}+\lambda
u_{zz}+6uu_z+\frac{1}{2}(6\lambda+5\beta)u^2-cu+K_1=0~,
\end{equation}
where $K_1$ is the integration constant. Moreover, employing
$u=w-\delta$, where the constant $\delta$ is equal to $\frac{-c\pm
\sqrt{c^2-4\alpha_1 K_1}}{2\alpha_1}$,
$\alpha_1=\frac{1}{2}(6\lambda+5\beta)$, and denoting
$\alpha_2=2\delta \alpha_1 + c$, one can get
%..................
\begin{equation}\label{can000}
w_{zzz}+\lambda w_{zz}+6(w-\delta)w_z+ w(\alpha_1 w-\alpha_2)=0~.
\end{equation}
Eq. (\ref{can000}) can be factorized in the form
\begin{equation}\label{can05}
[D_z-\phi _1(w)w_z-\phi _2(w)][D_{zz}-\phi _3(w)D_z-\phi _4(w)]w=0
\end{equation}
by introducing the appropriate $\phi_i$ functions. It is easily
shown by direct calculation that the factorization
\begin{equation}\label{can07}
\left[D_z+\frac{\alpha_1}{3}\right]\left[D_{zz}+(\lambda-\frac{\alpha_1}{3})D_z+\frac{3}{\alpha_1}(\alpha_1
w-\alpha_2)\right]w=0~,
\end{equation}
is allowed under the restriction
\begin{equation}\label{can07a}
\alpha_1^3-3\alpha_1^2\lambda-54\delta\alpha_1+27\alpha_2=0~.
\end{equation}
Therefore, the travelling wave solutions obtained for Eq.
(\ref{can000}) correspond to the case in which Eq. (\ref{can07a})
for the coefficients $\alpha_1$ and $\alpha_2$ is satisfied.
Furthermore, this restriction leads to
$c=-\frac{5}{6}\beta(\lambda+\frac{5}{6}\beta)^2$ which represents
the velocity of the travelling waves.

Let us consider now the extended factorization scheme \cite{wl08}.
Assuming
\begin{equation}\label{can07b}
\left[D_z^2+(\lambda-\frac{\alpha_1}{3})D_z+\frac{3}{\alpha_1}(\alpha_1
w-\alpha_2)\right]w=\Omega~,
\end{equation}
then Eq. (\ref{can07}) can be rewritten as the following system
%..........................
\begin{eqnarray}%\label{bsol1}
\Omega _z+\frac{\alpha_1}{3}\Omega=0,\label{can07c}\\
w_{zz}-\frac{5}{6}\beta w_z+\frac{3}{\alpha_1}(\alpha_1
w-\alpha_2)w=\Omega~.\label{can07d}
\end{eqnarray}
where $(\lambda-\frac{\alpha_1}{3})$ has been replaced with
$-\frac{5}{6}\beta$ in Eq. (\ref{can07d}). The first equation
implies $\Omega(z)=c_1e^{-\frac{\alpha_1}{3}z}$, where $c_1$ is an
integration constant. Thus, we can consider the solutions of the
second equation of the system, i.e.,
%..................
\begin{equation}\label{can08}
w_{zz}-\frac{5}{6}\beta w_z+3w^2-3\frac{\alpha _2}{\alpha
_1}w=c_1e^{-\frac{\alpha_1}{3}z}
\end{equation}
The transformations
$$
w=\lambda(z)W+\mu(z)~, \quad Z=\Phi(z)
$$
where
$$
\lambda(z)=(-2)^{1/5}e^{\frac{\beta}{3}z}~,\quad
\mu(z)=\left(\frac{\beta}{6}\right)^2+\frac{\alpha_2}{2\alpha_1}~,
\quad
\Phi(z)=\left(-\frac{1}{2}\right)^{2/5}\frac{6}{\beta}e^{\frac{\beta}{6}z}
$$
lead to the following canonical equation \cite{ince}
%..........................
\begin{equation}\label{can1}
\frac{d^2W}{dZ^2}=6W^2+S(z)
\end{equation}
where
$$
S(z)=-3\mu^2+3\frac{\alpha_2}{\alpha_1}\mu +
c_1e^{-\frac{\alpha_1}{3}z}~,
$$
which according to Ince's texbook has solutions free of movable
singularities other than poles only if $c_1=0$.
\medskip
However, this condition leads to $\mu=0$ and therefore to the
Painlev\'e case
%..........................
\begin{equation}\label{can1p}
\frac{d^2W}{dZ^2}=6W^2~,
\end{equation}
for which the solutions are
%..........................
\begin{equation}\label{can1p-sol}
W=C^2\left[\frac{-k^2}{1+k^2}+{\rm sn}^{-2}(CZ,k)\right]~.
\end{equation}
%..............
On the other hand, let us consider, for the same case $c_1=0$, the
second-order nonlinear differential equation coming out from the
factorization procedure
%..........................
\begin{equation}\label{can10}
\frac{d^2w}{dz^2}-\frac{5}{6}\beta\frac{dw}{dz}+3w\left[w-\alpha_3\right]=0
\end{equation}
where
$\alpha_3=\frac{\alpha_2}{\alpha_1}=\frac{\pm\sqrt{c^2-4\alpha_1K_1}}{3\lambda+\frac{5}{2}\beta}$.
We are now able to apply the factorization procedure introduced by
Rosu and Cornejo-P\'erez \cite{rc05} as a second-step procedure.
Eq. (\ref{can10}) can be factorized in the form
\begin{equation}
\left[D_z-f_2(w)\right]\left[D_z- f_1(w)\right]w=0~,\label{can10a}
\end{equation}
under the conditions
$$
\bigg\{
\begin{array}{cc}
f_1+f_2+\frac{df_1}{dw}w=& \frac{5}{6}\beta\\
f_1f_2w\qquad \quad=&3w(w-\alpha _3).\\
\end{array}
$$
Let $f_1=\sqrt{3}a_2(w^{1/2}-\alpha _{3}^{1/2})$ and
$f_2=\sqrt{3}a_{2}^{-1}(w^{1/2}+\alpha _{3}^{1/2})$. From the last
factorization conditions one can get after some algebra the
following values of the parameters $a_2=\pm i \sqrt{\frac{2}{3}}$
and $\alpha_3^{1/2}=\pm i \frac{\sqrt{2}}{6}\beta$ for which this
type of factorization is possible. Therefore, solutions of
$$
[D-f_1]w=0
$$
will be solutions of the factorized equation as well. The latter
equation has the explicit form
%..........................
\begin{equation}\label{can11}
\frac{dw}{dz}\pm i\sqrt{2}w^{3/2}\pm i\sqrt{2}\alpha
_{3}^{1/2}w=0~.
\end{equation}
The solutions of these two Bernoulli equations can be directly
written down. Taking into account that $u=w-\delta$ we immediately
get:
%..........................
\begin{eqnarray}%\label{bsol1}
u_{1,2}=\left[-\frac{3\sqrt{2}}{\beta}i+e^{\frac{\beta}{6}(z-z_0)}\right]^{-2}-\frac{5}{36}\lambda
\beta
-\frac{\beta^2}{36}\left[\frac{25}{6}\pm1 \right]\\
u_{3,4}=\left[\frac{3\sqrt{2}}{\beta}i+e^{-\frac{\beta}{6}(z-z_0)}\right]^{-2}-\frac{5}{36}\lambda
\beta -\frac{\beta^2}{36}\left[\frac{25}{6}\pm1 \right]\\
\end{eqnarray}

Choosing now the factorization functions
$f_1=\sqrt{3}a_2(w^{1/2}+\alpha _{3}^{1/2})$ and
$f_2=\sqrt{3}a_{2}^{-1}(w^{1/2}-\alpha _{3}^{1/2})$ the following
two Bernoulli equations are obtained
%..........................
\begin{equation}\label{can22}
\frac{dw}{dz}\mp i\sqrt{2}w^{3/2}\pm i\sqrt{2}\alpha
_{3}^{1/2}w=0~,
\end{equation}
whose solutions are
\begin{eqnarray}
u_{5,6}=\left[\frac{3\sqrt{2}}{\beta}i+e^{\frac{\beta}{6}(z-z_0)}\right]^{-2}-\frac{5}{36}\lambda
\beta -\frac{\beta^2}{36}\left[\frac{25}{6}\pm1 \right]\\
u_{7,8}=\left[-\frac{3\sqrt{2}}{\beta}i+e^{-\frac{\beta}{6}(z-z_0)}\right]^{-2}-\frac{5}{36}\lambda
\beta -\frac{\beta^2}{36}\left[\frac{25}{6}\pm1 \right]~.\\
\end{eqnarray}
%\frac{c\pm \sqrt{c^2-4\alpha _1K_1}}{2\alpha_1}

\bigskip
On the other hand, combining the factorization conditions
$$
f_1f_2w=F(w)~, \qquad
f_2+\frac{d}{dw}\left(f_1w\right)=\frac{5}{6}
$$
and introducing the function $l(w)=f_1(w)w$ one obtains an Abel
equation of the form
%..........................
\begin{equation}\label{abel1}
l\frac{dl}{dw}-\frac{5}{6}\beta l=-3w^2+3\alpha _3w~.
\end{equation}
The solution of this equation can be written as follows
\cite{polyanin}
\begin{equation}\label{abel-sol1}
w=\frac{1}{2}\left(\frac{\beta}{3}\right)^2e^{\frac{\beta}{3}(z-z_0)}{\cal
P}\left(e^{\frac{\beta}{3}(z-z_0)}+c_2,0,1\right)~,
\end{equation}
which is expressed in terms of the Weierstrass ${\cal P}$
function. Therefore,
%..........................
\begin{equation}\label{abel-sol}
u(z)=\frac{1}{2}\left(\frac{\beta}{3}\right)^2e^{\frac{\beta}{3}(z-z_0)}{\cal
P}\left(e^{\frac{\beta}{3}(z-z_0)}+c_2,0,1\right)
-\frac{5}{36}\lambda \beta
-\left(\frac{\beta}{6}\right)^2\left[\frac{25}{6}\pm1 \right]~.
\end{equation}
We notice that the latter solution although similar in form to a
solution mentioned by Porubov \cite{p93} is different by an
additive constant that depends on the coefficients of the PKdV
equation and by the variable of the Weierstrass ${\cal P}$
function which in the case of Porubov's result is
$\exp(y)=\exp(e^{\gamma (z-z_0)})$, $\gamma =$ constant.

\medskip

The Weierstrass component of the solution (\ref{abel-sol}) can be
also written in the following form, see also \cite{est06}
%..........................
\begin{equation}\label{abel-sol2}
\frac{1}{2}\left(\frac{\beta}{3}\right)^2
\frac{e^{\frac{\beta}{3}z}}{4\sqrt{3}}k_0^2 \left[
1+\sqrt{3}\frac{1+{\rm cn}\left(k_0(e^{\frac{\beta}{6}z}+c_2
e^{\frac{\beta}{6}z_0})|m\right)}{1-{\rm
cn}\left(k_0(e^{\frac{\beta}{6}z}+c_2
e^{\frac{\beta}{6}z_0})|m\right)}\right]~,
\end{equation}
%.........................
where $k_0=2H_{2}^{1/2}e^{-\frac{\beta}{6}z_0}$,
$H_2=\frac{\sqrt{3}}{4^{1/3}}$,
$m=\frac{1}{2}-\frac{\sqrt{3}}{4}$. Expanding now the cnoidal
function in a Taylor series
$$
{\rm cn}\left(k_0(e^{\frac{\beta}{6}z}+c_2
e^{\frac{\beta}{6}z_0})|m\right)=1-\frac{1}{2}k_0^2\left(e^{\frac{\beta}{6}z}+c_2
e^{\frac{\beta}{6}z_0}\right)^2+...~,
$$
one gets in the small $k_0$ limit:
$$
\lim _{k_0\rightarrow
0}u(z)=\frac{1}{2}\left(\frac{\beta}{3}\right)^2\left(\frac{e^{\frac{\beta}{6}(z-z_0)}}{c_2+e^{\frac{\beta}{6}(z-z_0)}}\right)^2+
{\rm const}~. =
\left(\frac{3\sqrt{2}}{\beta}+c_3e^{\frac{\beta}{6}(z-z_0)}\right)^{-2}-\frac{5}{36}\lambda
\beta-\left(\frac{\beta}{6}\right)^2\left[\frac{25}{6}\pm
1\right]~,
$$
which is a simple particular real PKdV solution.

%
%\begin{eqnarray} % EQUATION 14
%{\rm z_{2}^{-k_2}w_{2}^{+}(t)={\cal
%A}\,z_{1}^{k_1} %z_{2}^{(q-\frac{1}{2})}
%\,_{2} F_{1}\left[k_1+k_2+1,k_1+k_2,1+2k_1\,;-\frac{z_{1}}{2}\right]}\nonumber\\
%- {\rm {\cal B}\,e^{-i(1+2k_1)\pi}\left(\frac{4}{z_1}\right)^{k_1}
%%%%%z_{2}^{(q-\frac{1}{2})}
%\, _{2}F_{1}\left[-k_1+k_2,
%-k_1+k_2+1,1-2k_1\,;-\frac{z_{1}}{2}\right]} %e^{i(1-2p)\pi}
%\end{eqnarray}

%\newpage
\bigskip
%%44444444444444444444444444444444444444444444444444444444444444444444444444444444444
%%%%%%%%%%%%%%%%%%%%
%%%%%%%%%%%%%%%%%%%%%%
%\section{Conclusion}
%\underline{{\em Interpretation}}

%\medskip

\noindent In conclusion, after performing the travelling variable reduction for the PKdV equation we have jointly used recent
factorization methods to obtain some new exact travelling wave solutions of this equation in the particular case when the coefficients fulfill the condition (\ref{can07a}). This is equivalent to saying that the factorization of the ODE travelling form of PKdV equation can be performed only for a particular value of the velocity parameter and leads to a second order differential equation that has the Painlev\'e property, a fact that pinpoints the connection between the technique of factorizations and the Painlev\'e analysis. The latter connection has been already noticed by Gilson and Pickering for other types of nonlinear third-order partial differential equations \cite{gp95}.

%\nonumsection{Acknowledgements}

%\noindent
%This work was supported by a Project from CONACyT (No.~).

\end{document}